\documentclass[preprint,nofootinbib]{revtex4}
\usepackage{graphicx}
\usepackage{amsmath,amssymb}
\usepackage[colorlinks,citecolor=blue,linkcolor=blue,urlcolor=blue]{hyperref}
\usepackage[babel]{csquotes}
\usepackage{mathrsfs}
\usepackage{enumerate}
\usepackage{caption}
\usepackage{ragged2e}
\usepackage{hyperref}
\usepackage[margin=0pt,textfont={small,it},labelfont={small,bf},
labelsep=colon,hypcap=true,justification=RaggedRight]{caption}
\usepackage{enumitem}

\def\be{\begin{equation}}
\def\ee{\end{equation}}

\begin{document}
\title{Spin fluctuations and black hole singularities:\\ the onset of quantum gravity is spacelike}

\author{Eugenio Bianchi }
\email{ebianchi@gravity.psu.edu}
\affiliation{\footnotesize Institute for Gravitation \& the Cosmos and Department of Physics, The Pennsylvania State University, University Park, PA 16802, USA}
 
\author{Hal M. Haggard}
\email{haggard@bard.edu}
\affiliation{\footnotesize Physics Program, Bard College, 30 Campus Road, Annandale-on-Hudson, NY 12504, USA,\\
Perimeter Institute, 31 Caroline Street North, Waterloo, ON, N2L 2Y5, CAN}

\author{${}$}

\begin{abstract}
Due to quantum fluctuations, a black hole of mass $M$ represents an average over an ensemble of black hole geometries with angular momentum.  This observation is apparently at odds with the fact that the curvature singularity inside a rotating black hole is timelike, while the one inside a non-rotating black hole is spacelike. Is the average of timelike singularities really spacelike? We use the Bekenstein-Hawking entropy formula to introduce a microcanonical ensemble for spin fluctuations and show that the onset of quantum gravity is always spacelike. We discuss the impact of this result on singularity resolution in quantum gravity and hint at the possibility of an observational test.  
\end{abstract}

\flushbottom
\maketitle

\section{Introduction: the puzzle}
In the quantum gravitational treatment of a spherically symmetric black hole a puzzle arises. The symmetric Schwarzschild black hole has a singularity at its core where the curvature diverges and general relativity breaks down. This singularity is sometimes described as a moment of time, or as the end of spacetime, because all observers that enter the black hole horizon reach it in a finite time no matter where or how they enter. This structure is usefully summarized in a Penrose causal diagram where a horizontal jagged line represents the spacelike singularity. 
However, in quantum gravity, a classical spacetime like Schwarzschild is actually a mixture of microstates in an ensemble of quantum spacetimes; the ensemble characterizes the quantum fluctuations of the spacetime state and the number of microstates in the ensemble is given by the exponential of the Bekenstein-Hawking entropy. In particular, the members of this ensemble will carry angular momentum and can be organized as a sum over spins $\vec{J}$,
\begin{equation}
   \includegraphics[width=.7\textwidth]{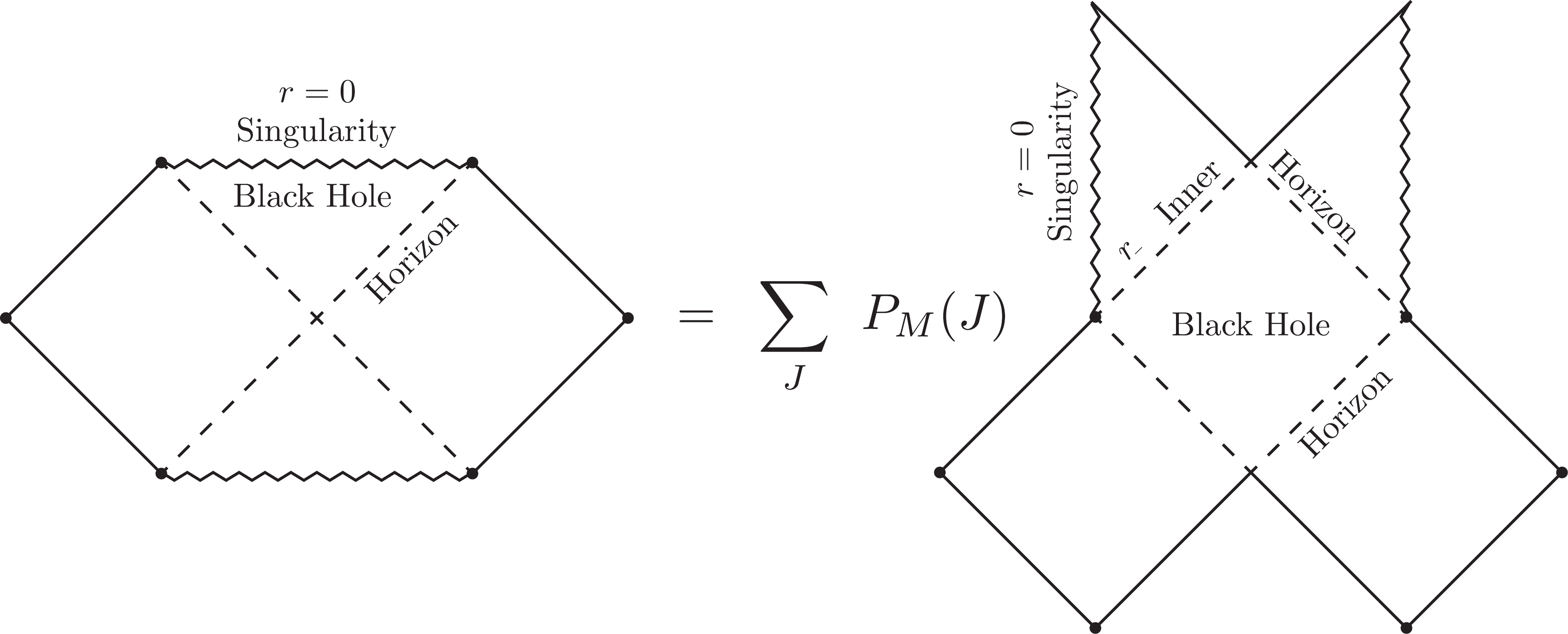} \qquad,
\label{eq:Penroses}
\end{equation}
where on the right the sum is over spin magnitude $J = |\vec{J}|$ and $P_M(J)$ is the weight for a black hole of mass $M$ and spin $J$. A rotating black hole is described by a Kerr metric and its singularity is timelike as shown in its causal diagram, the vertical jagged line of Eq.~(\ref{eq:Penroses}). It is because of this that a puzzle arises. We broadly agree that quantum gravity resolves all of these singularities, but where or when should we locate the onset of quantum gravity? What is the behavior of the quantum gravitational takeover? Is it spacelike, like the singularity of Schwarzschild, or timelike, like that of Kerr?  And how is it that the spacelike onset of quantum gravity for a spherically symmetric black hole arises from the mixture of a set of rotating quantum black holes as described by Eq.~(\ref{eq:Penroses})? 

These questions about spin fluctuations and black hole interiors lead to a fundamental recognition about the onset of quantum gravity inside a black hole: despite spin fluctuations, the onset of quantum gravity is spacelike. There are two assumptions that go into our proof that spin fluctuations do not change the nature of the onset of quantum gravity: 
\begin{enumerate}
\item[(i)] that quantum black holes fluctuate according to a probability distribution determined by the Bekenstein-Hawking entropy, and
\item[(ii)] that Planckian values for any curvature invariant lead to quantum gravity. 
\end{enumerate}
These two assumptions are compatible with all current approaches to quantum gravity. We describe in more detail the two assumptions, offer several elaborations on them, and discuss their consequences.

\section{Bekenstein-Hawking entropy and the black-hole spin ensemble}
A black hole of mass $M$ and spin $\vec{J}$ has an entropy proportional to its horizon area $A(M,J)$ and given by the Bekenstein-Hawking formula \cite{Bekenstein},
\begin{equation}
S_{\mathrm{\emph{BH}}}(M,J)=\frac{A(M,J)}{4\ell_{P}^{2}}=\Bigg(1+\sqrt {1-\bigg(\frac{J}{GM^{2}/c}\bigg)^{2}\;}\; \Bigg) \frac {2\pi M^{2}} {m_{P}^{2}}\,,
\label{eq:}
\end{equation}
where $\ell_P \equiv \sqrt{\hbar G/c^3}$ is the Planck length and $m_P \equiv \sqrt{\hbar c/G}$ the Planck mass. The statistical mechanical interpretation of this entropy is that a black hole of mass $M$ and spin $\vec{J}$ is a mixture of $\mathcal{N}\sim \exp S_{\mathrm{\emph{BH}}}(M,J)$ microstates.  We show that an immediate consequence of the presence of these microstates is the existence of a black-hole spin ensemble: the probability that a black hole of mass $M$ is found to have spin $J$ is given by the probability distribution
\begin{equation}
P_{M}(J)=\frac{\displaystyle e^{\,A(M,\, J)/4\ell_{P}^{2}}\;J^{2}}{\displaystyle \int_{0}^{J_{\text{max}}}\;e^{\,A(M,\, J)/4\ell_{P}^{2}}\;J^{2}\,d J}\,,
\label{eq:PMJ}
\end{equation}
where $J_{\text{max}}=GM^{2}/c$. We will derive this formula and further clarify the nature of the spin ensemble.

In asymptotically flat spacetimes, the ADM mass and angular momentum identify irreducible representations of the asymptotic Poincar\'e group \cite{Arnowitt:1962hi}. In a quantum theory of gravity with asymptotically flat boundary conditions, the mass and angular momentum
\begin{equation}
\hat{M}\,,\quad \hat{J},
\label{eq:}
\end{equation}
provide a commuting set of operators whose eigenspaces are labeled by representations of the asymptotic Poincar\'e group. We denote simultaneous eigenstates by $|M,j;\alpha\rangle $ with
\begin{align}
\hat{M}\,|M,j;\alpha\rangle&=M\,|M,j;\alpha\rangle\, , \quad \text{and} \\[.5em]
\hat{J}\;|M,j;\alpha\rangle&=\hbar\, \sqrt{j(j+1)}\;|M,j;\alpha\rangle\,.
\label{eq:}
\end{align}
For massive states, $M>0$, the spin is labeled by the half-integer $j$. The index $\alpha$ labels the states of an orthonormal basis for the Hilbert space. The Bekenstein-Hawking formula for black hole entropy indicates that the vast majority of the states $|M,j;\alpha\rangle$ are black-hole microstates.

The Hilbert space of the quantum theory decomposes in sectors of given mass and spin,
\begin{equation}
\mathcal{H}=\bigoplus_{M}\mathcal{H}_{M}\,,\quad\text{with}\quad\mathcal{H}_{M}=\bigoplus_{j=0,\frac{1}{2},1,\ldots}\mathcal{H}_{Mj}\,.
\label{eq:}
\end{equation}
A rotating black hole of mass $M$ and spin $J=\hbar\sqrt{j(j+1)}$ is understood as the maximally-mixed state at fixed energy and angular momentum and is given by the density matrix
\begin{equation}
\rho(\:\parbox[c]{2em}{\includegraphics[width=2em]{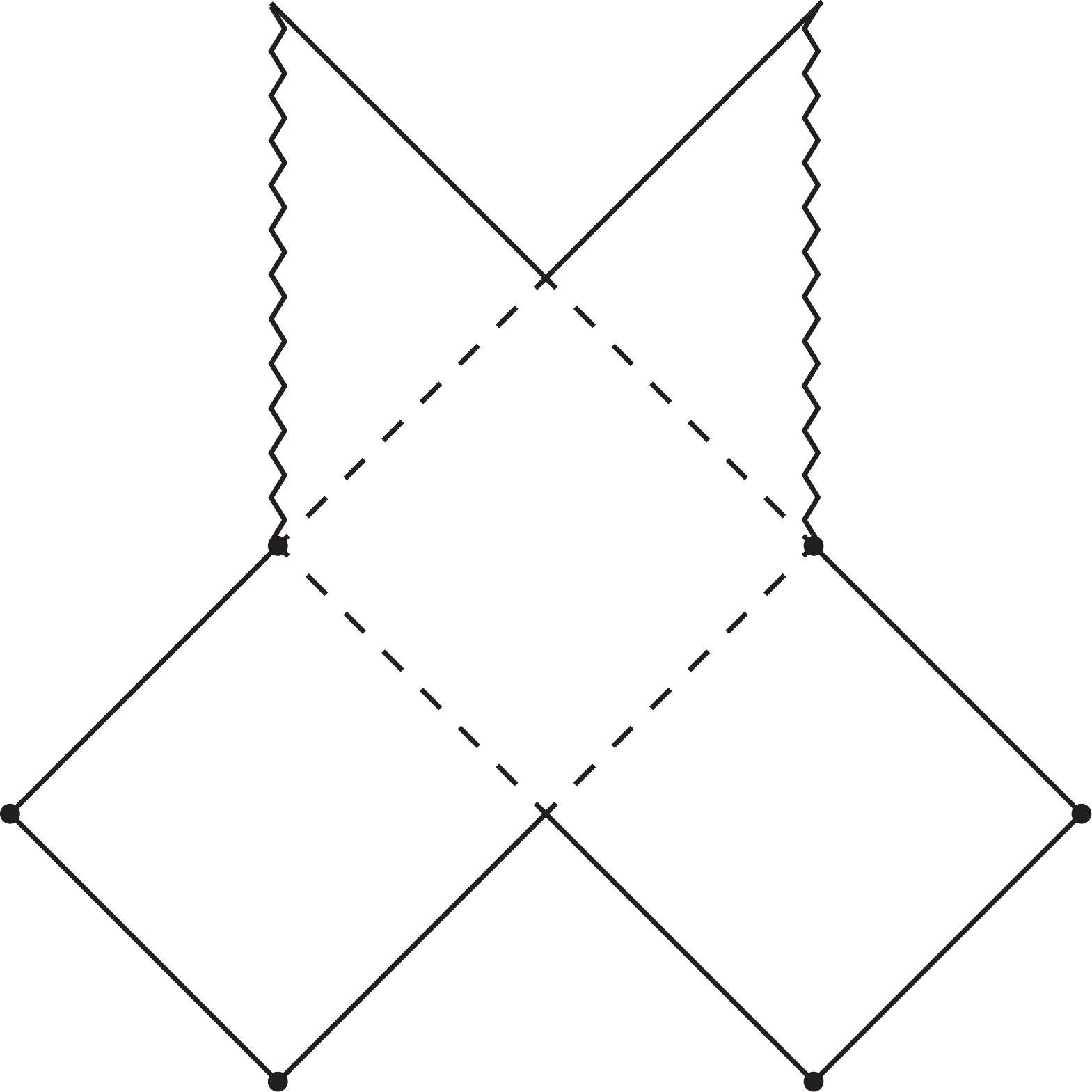}}\;)\equiv\rho_{Mj}=\frac{1}{\dim \mathcal{H}_{Mj}}\!\!\sum_{\alpha=1}^{\dim \mathcal{H}_{Mj}}\!\!|M,j;\alpha\rangle\langle M,j;\alpha|\,.
\label{eq:}
\end{equation}
The dimension of the Hilbert space $\mathcal{H}_{Mj}$ can be computed via semiclassical methods starting from the canonical partition function written in terms of the Euclidean action, $Z(\beta,\vec{\omega})=\text{Tr}\big(e^{-\beta \hat{M}-\vec{\omega}\cdot \vec{J}/\hbar}\,\big) \,=\, \int[Dg_{\mu\nu}] \,e^{-S_{E}[g_{\mu\nu}]/\hbar}$, \cite{Gibbons:1976ue,Brown:1992bq,Sen:2012dw}. One finds that, up to logarithmic corrections, the entropy of the mixture of microstates is given by the Bekenstein-Hawking formula,
\begin{equation}
S(\:\parbox[c]{2em}{\includegraphics[width=2em]{kerr.pdf}}\;)=-\text{Tr}(\rho_{Mj}\log \rho_{Mj})\;=\;\log(\dim \mathcal{H}_{Mj})\;=\;\textstyle\frac{A(M,J)}{4\ell_{P}^{2}}\;+\;O\big(\log \frac{M}{m_{P}}\big)\,.
\label{eq:Skerr}
\end{equation}
On the other hand, a black hole of mass $M$ is understood as given by the \emph{microcanonical ensemble} in which only the mass $M$ is fixed. The state is given by the density matrix
\begin{equation}
\rho(\:\parbox[c]{2.5em}{\includegraphics[width=2.5em]{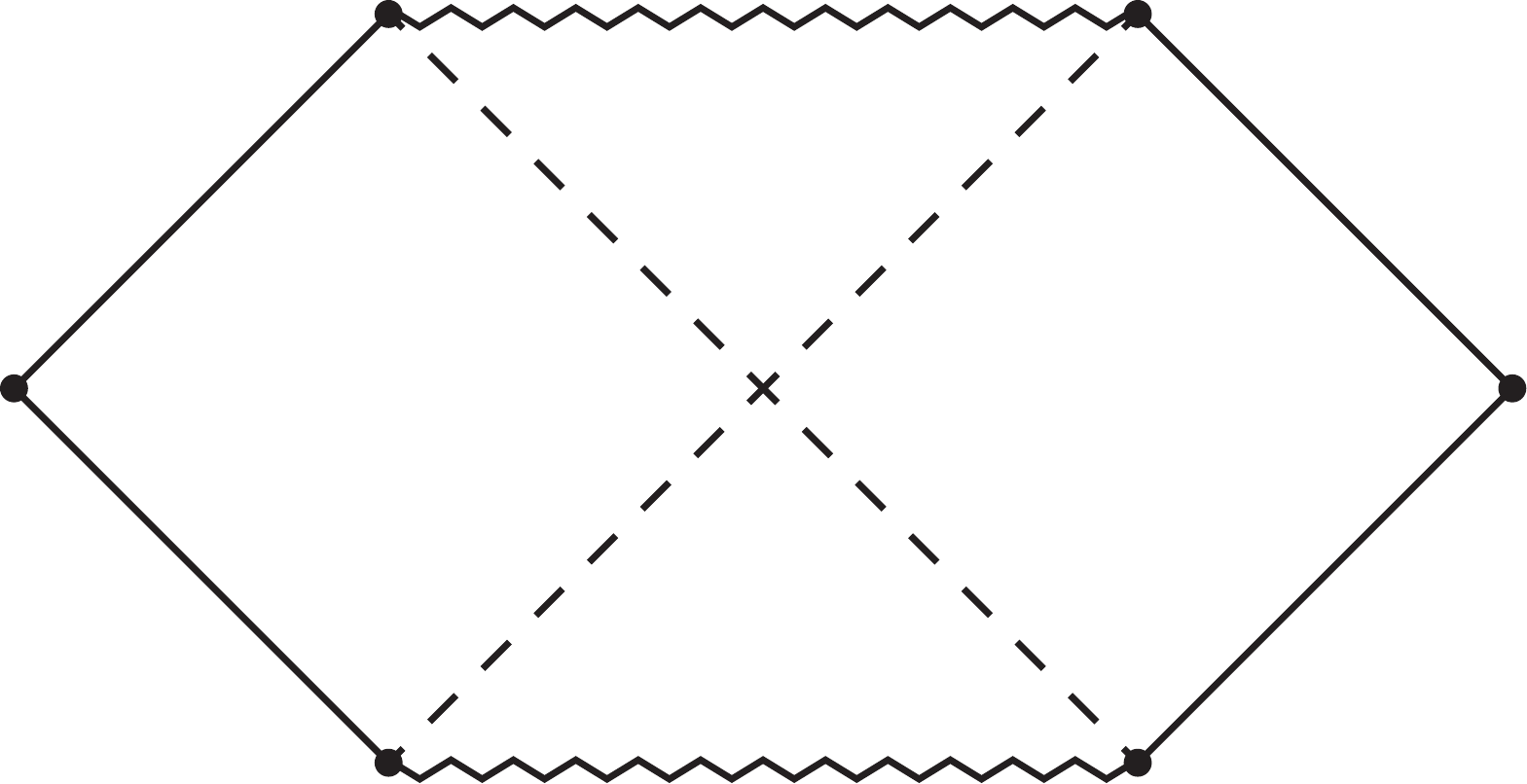}}\:)\equiv\,\rho_{M}=\frac{1}{\dim \mathcal{H}_{M}}\sum_{j}\!\!\sum_{\alpha=1}^{\dim \mathcal{H}_{Mj}\!\!}\!\!\!|M,j;\alpha\rangle\langle M,j;\alpha|\;,
\label{eq:microcanonical}
\end{equation}
where the dimension $\dim \mathcal{H}_{M}=\sum_{j} \dim \mathcal{H}_{Mj}$ can again be computed via semiclassical methods. In fact, taking into account the logarithmic corrections $\log\!\big(\!\dim \mathcal{H}_{Mj}\big)\;=\;\frac{A(M,J)}{4\ell_{P}^{2}}\;+\frac{1}{2}\big(c_{0}-\frac{3}{2}\big)\,\log \frac{A(M,J)}{4\ell_{P}^{2}}+\ldots\,$ in Eq.~(\ref{eq:Skerr}) (with $c_{0}=\frac{212}{45}$ determined by graviton loop corrections and the $-\frac{3}{2}$ term due to the choice of ensemble with fixed spin \cite{Sen:2012dw}), one finds that the entropy of the microcanonical state defining a Schwarzschild black hole is
\begin{equation}
S(\:\parbox[c]{2em}{\includegraphics[width=2em]{schwarzschild.pdf}}\;)=-\text{Tr}(\rho_{M}\log \rho_{M})\;=\;\log(\dim \mathcal{H}_{M})\;=\;\textstyle\frac{A(M,0)}{4\ell_{P}^{2}}\;+\;c_{0}\,\log \frac{A(M,0)}{4\ell_{P}^{2}}+\,\cdots\;.
\label{eq:Sschw}
\end{equation}
Computing the logarithmic corrections explicitly highlights a remarkable property of the microcanonical ensemble. The logarithmic corrections show that the ensemble of states with $M$ fixed has a larger entropy than the ensemble with $M$ and $J$ fixed and with the spin set to $J=0$.  The microcanonical ensemble $\rho_M$ is not the zero-spin limit of the ensemble  $\rho_{Mj}$.

The microcanonical ensemble (\ref{eq:microcanonical}) can be written as a mixture of states with fixed mass and spin as $\rho_{M}=\sum_{j}p_{n}(j)\,\rho_{Mj}$, where the probability of spin $j$ is given by 
\begin{equation}
p_{M}(j)=\frac{\dim \mathcal{H}_{Mj}}{\sum_{j} \dim \mathcal{H}_{Mj}}\,.
\label{eq:pnj}
\end{equation}
This is the ratio of dimensions of the two Hilbert spaces, the one at fixed mass and spin, and the one at fixed mass. The relation between the microcanonical state of a black hole of mass $M$ and the one of a black hole with fixed mass and spin is then
\begin{equation}
\rho(\:\parbox[c]{2.5em}{\includegraphics[width=2.5em]{schwarzschild.pdf}}\:)=\sum_{j}\,p_{M}(j)\;\rho(\:\parbox[c]{2em}{\includegraphics[width=2em]{kerr.pdf}}\;)\,.
\label{eq:rho=p(j)rho}
\end{equation}
For large mass $M\gg m_{P}$, the dimension of the Hilbert space $\mathcal{H}_{Mj}$ is well approximated by the leading-order semiclassical term, $\dim \mathcal{H}_{Mj}\,\sim e^{\,A(M,J)/4\ell_{P}^{2}}\delta M\;4\pi J^{2}\delta J$. In this limit, the probability $p_{M}(j)$ reduces to the formula (\ref{eq:PMJ}) for $P_{M}(J)$, and the expression (\ref{eq:rho=p(j)rho}) reproduces the diagrammatic formula, Eq.~(\ref{eq:Penroses}).

\section{The onset of quantum gravity} 
No matter the specific theory of quantum gravity, the common expectation is that quantum gravitational phenomena are significant at Planckian curvatures. To establish the claim that---despite spin fluctuations---the onset of quantum gravity is spacelike, we consider a curvature invariant and discuss its behavior for the Kerr metric.

The Kerr metric is  determined by two length scales, the Schwarzschild radius $r_S  \equiv 2GM/c^2$, and the Kerr parameter $a \equiv J/Mc$ where $M$ is the mass and $J$ is the angular momentum of the black hole. As this metric is a vacuum solution of the Einstein equations, the interesting curvature invariants only start with contractions of the full Riemann tensor, or equivalently the Weyl tensor $C_{\mu \nu \rho \sigma}$. Intriguingly, the two independent quadratic invariants formed from the Weyl tensor can be collected into a single complex Weyl scalar $W \equiv  \left( C_{\mu \nu \rho \sigma}+i\ {}^*C_{\mu \nu \rho \sigma} \right)C^{\mu \nu \rho \sigma}$, 
where ${}^*C_{\mu \nu \rho \sigma} \equiv \frac{1}{2} \epsilon^{\phantom{\rho \sigma} \alpha \beta}_{\rho \sigma} C_{\mu \nu \alpha \beta}$ is the dual Weyl tensor. In Boyer-Lindquist coordinates this invariant has a remarkably simple magnitude
\begin{equation}
|W(r,\theta)| =  \sqrt{\left( C_{\mu \nu \rho \sigma}C^{\mu \nu \rho \sigma}\right)^2+\left({}^*C_{\mu \nu \rho \sigma} C^{\mu \nu \rho \sigma} \right)^2} = \frac{12\, r_S^2}{(r^2 + a^2 \cos^2 \theta)^3}.
\end{equation}
While the real and imaginary parts of $W$ generally oscillate as you approach the ring singularity of Kerr, this invariant does not and it has a transparent dependence on the angle of approach. For $a=0$, it is the Kretschmann invariant.

\begin{figure}[h!] 
\begin{center}
  \includegraphics[width=3.5in]{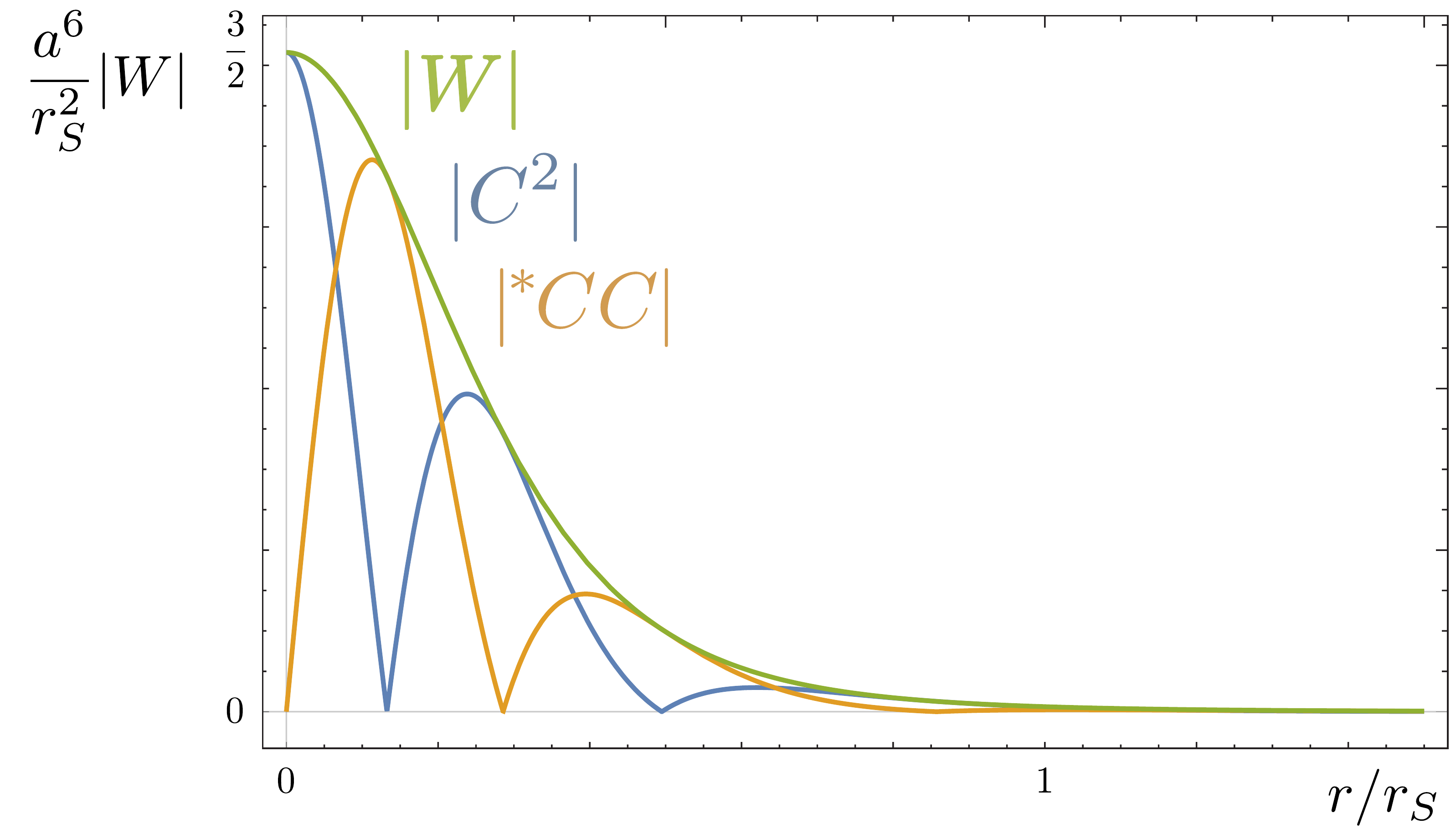} 
\caption{ The Weyl scalar $|W|$ is a useful measure of the strength of curvature because it envelopes both the standard vacuum Kretschman $|C^2|$ and Euler-Pontryagin $|{}^* CC|$ scalars. All three curves are evaluated at a polar angle of $\theta=\pi/4$, for a Kerr parameter $a=0.7$, and for a black hole mass with equivalent Schwarzschild radius of $r_S =1$ km. On the equatorial plane $\theta=\pi/2$ the curvature diverges near zero radius.}
   \label{Small}
\end{center}
\end{figure}

For a vacuum Kerr spacetime we define the onset of the quantum gravity region as the surface on which the curvature invariant $|W|$ first becomes Planckian, or equivalently,
\begin{equation}
\text{the Planckian curvature radius $r_{*}$ is defined by} \quad |W(r_{*},\pi/2)| = 1/\ell_{P}^4\ .
\end{equation} 
There is a candidate mechanism that can protect the onset of quantum gravity against the odd discontinuous flip of the singularity from spacelike to timelike under a spin fluctuation: the generation of an inner horizon. The horizon condition $\Delta \equiv r^2 - r_S r+a^2=0$ for a Kerr black hole has two roots $r_{\pm} \equiv  {\displaystyle \frac{r_S}{2}} \left[ 1\pm\sqrt{1-\left({2a/r_S} \right)^2} \ \right]$. The outer root $r_+$ acts as the event horizon for the eternal black hole, but as a black hole spins up a second horizon separates from $r=0$ and develops at $r_-$. This inner horizon is null and is a Cauchy horizon beyond which the Cauchy development cannot be continued \cite{Frolov}. Considerable effort has been devoted to the study of the classical and the quantum stability of the Cauchy horizon \cite{Penrose:1968,PoissonIsrael:1989}, a topic we will take up in the Discussion section below. 

In the case of spin fluctuations of a black hole of mass $M$, it is possible that quantum gravity takes over before arriving at the inner horizon. If so, then the odd flip in the nature of the singularity at zero spin is an irrelevant mathematical structure that does not play a role in the physics of the black-hole spin ensemble. The onset of quantum gravity would be spacelike not only for the Schwarzschild term on the left-hand side of Eq.~(\ref{eq:Penroses}), but also for the Kerr terms on the right-hand side which should be drawn as in Figure~\ref{Small} {\bf (a)}.  

\begin{figure}[t] 
  \includegraphics[width=2.7in]{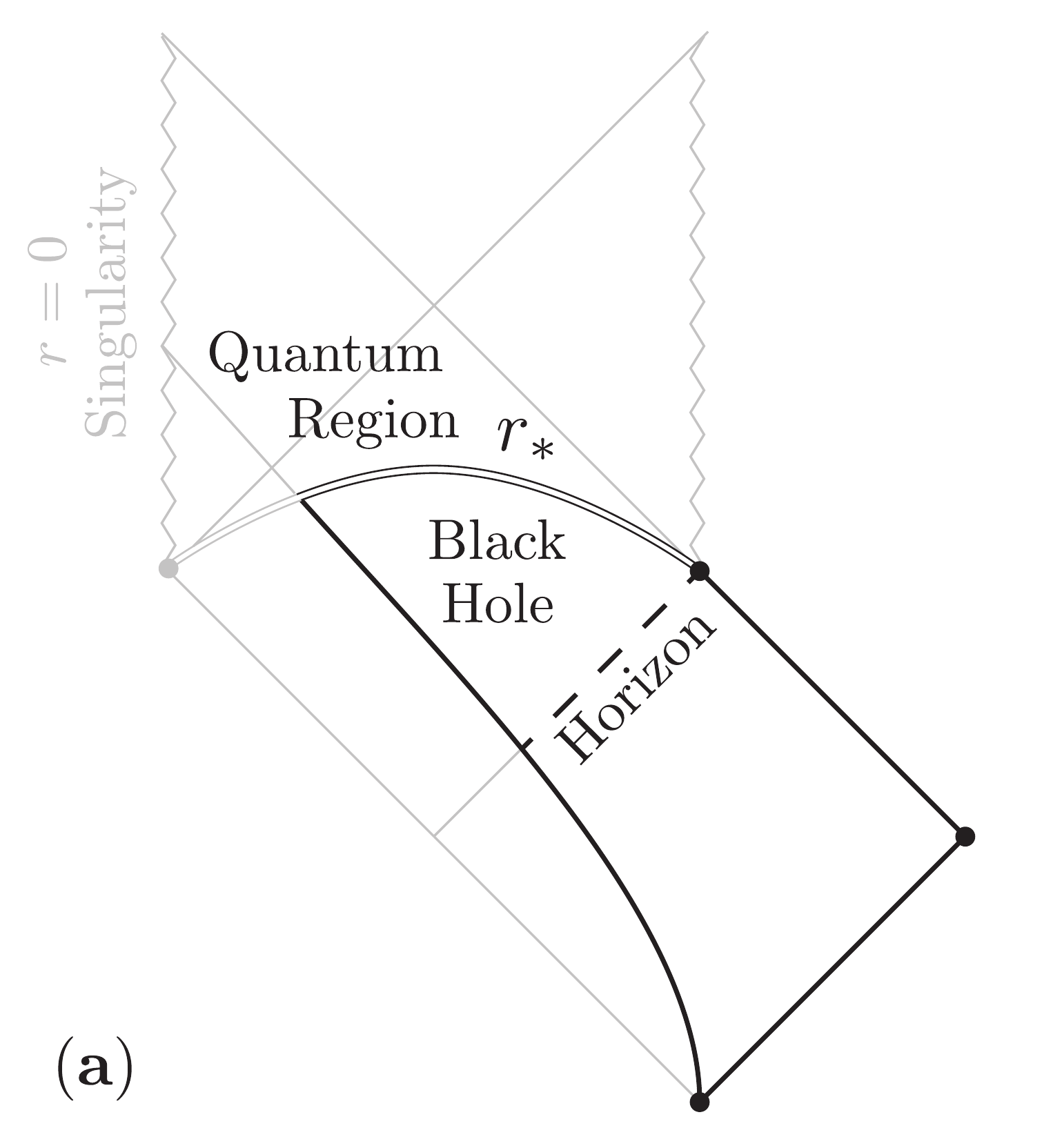}  \hspace{1.5cm}  \includegraphics[width=2.95in]{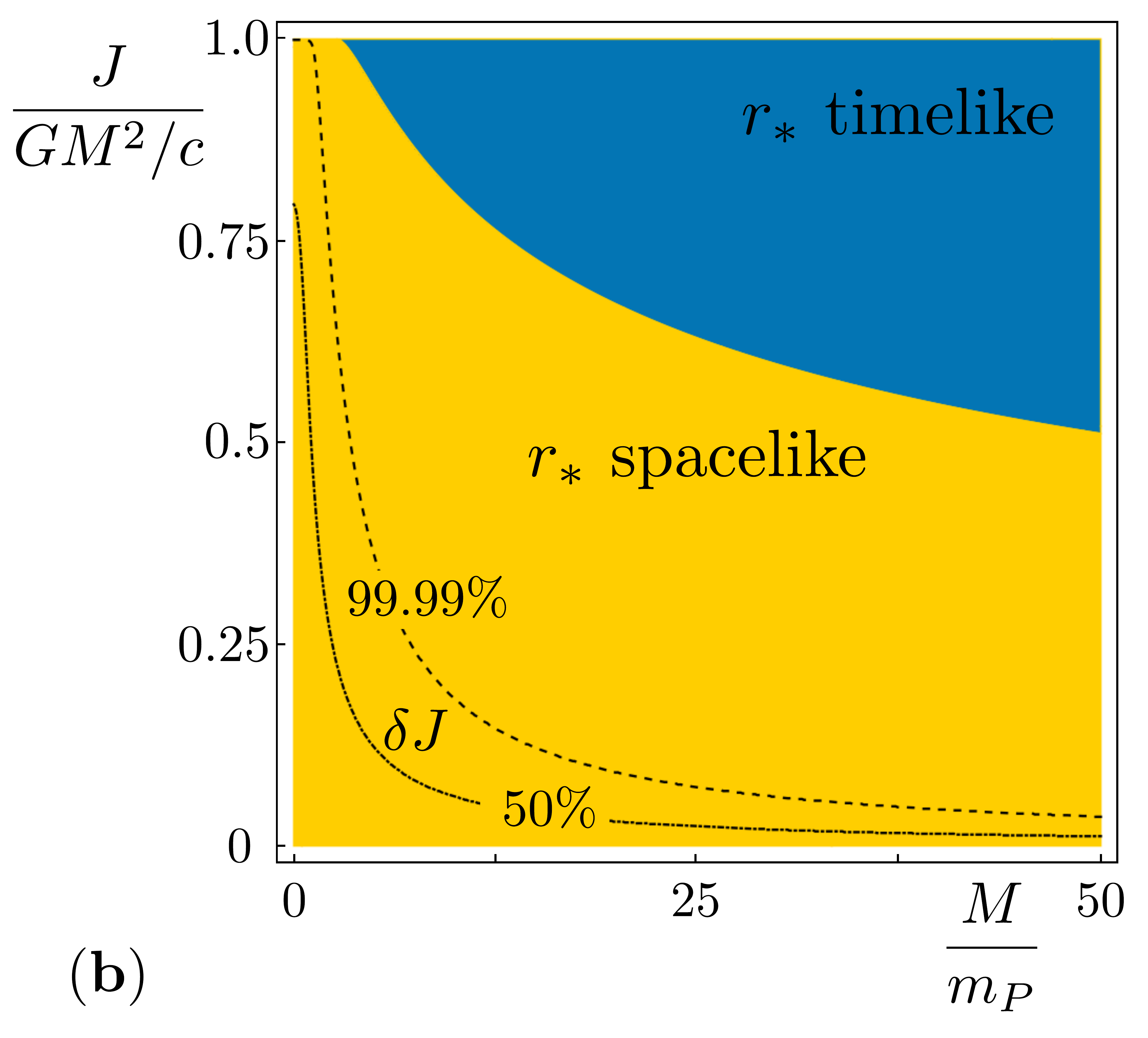}   
  
\caption{ {\bf (a)} The physical Penrose diagram of a Kerr black hole. For a quantum fluctuation $\delta J$ of the angular momentum, the timelike singularities lie deep in the quantum region and the onset of quantum gravity at $r_*$ is  spacelike (double line). {\bf (b)}  The dimensionless parameter space of a Kerr black hole. The regions where $r_*$ is spacelike (yellow) and timelike (blue) are shown as well as the contours below which  $99.99\%$ and $50\%$ of all spin fluctuations are contained. }
   \label{Small}
\end{figure}

To test this conjecture we characterize the black hole in its $J$ and $M$ parameter space,   Fig.~\ref{Small}~{\bf (b)}.  Computing black hole spin fluctuations requires quantum gravity and can be achieved using the black-hole spin ensemble (\ref{eq:PMJ}) derived from the Bekenstein-Hawking entropy formula. The fluctuation contours of Fig. \ref{Small}~{\bf (b)} are constructed by scanning through the masses and at each mass identifying the value of $J$ for which the indicated fraction of the distribution is achieved. In the limit $M \gg m_{P}$ the spin fluctuations is $ \delta J =\frac{ \hbar}{\sqrt{2\pi}} \frac{ M}{m_{P}} $. We compute the Planckian curvature radius $r_{*}$ and---using Eq.~(\ref{eq:PMJ})---the 50\% and 99.99\% intervals for the angular momentum fluctuations. The light (yellow) region of the plot shows the parameter space for which the onset of quantum gravity at $r_*$ is spacelike. The angular momenta resulting from quantum fluctuations (black contours) are always hidden in the quantum fog of high curvature and this resolves our puzzle. Thus, Figure~\ref{Small} {\bf (a)} depicts a physical version of the causal diagram of a rotating black hole with small spin. The onset of quantum gravity is spacelike and we have depicted it with a double line. Quantum fluctuations in a black hole's spin {\em do not change the character of the onset of quantum gravity}; it is always spacelike.

\vspace{-.08in}

\section{Discussion}

\vspace{-.08in}

The analysis of spin fluctuations of a quantum Schwarzschild black holes provides a new probe of the spacelike nature of the onset of quantum gravity. In this section we discuss the relation of our results to previous investigations of the onset of quantum gravity inside black holes and elaborate on the role played by the spin ensemble.

The angular momentum imparted to a black hole by a spin fluctuation is quite small, only $4000\; \text{kg}\: \text{m}^2\: \text{s}^{-1}$ for a solar mass black hole (which corresponds to a Planck-scale Kerr parameter $a$). Thus, for black holes with larger angular momenta it is no longer the case that the $r_*$ determined by the background curvature alone is spacelike and precedes the Cauchy horizon.   However, quite early on it was noted that the interaction between fields and the background might lead to large curvatures in the neighborhood of the Cauchy horizon. Penrose \cite{Penrose:1968} recognized that a blue shift instability at the Cauchy horizon indicates new physics. This was later convincingly established by Poisson and Israel who found that in a model of collapse to a rotating black hole massive outflow from the collapsing star leads to an exponentially growing mass parameter in the interior and to Planckian curvatures before the inner horizon \cite{PoissonIsrael:1989}, see also \cite{Ori:1991}. Indeed mass inflation, as they called it, is quite generic when a black hole accretes. Fo an accreting black hole, even weakly accreting inflows interact with the small outflow to drive the exponential growth \cite{HamiltonAvelino:2010}.  Other developments include a complementary analysis of the Cauchy horizon \cite{Flanagan:1997}, analysis of the ingoing inner horizon (the complement to the Cauchy horizon) \cite{MarolfOri:2012}, and the development of precise theorems \cite{DafermosLuk:2017}. All of this work points to singular behavior at the inner horizon, often due to back reaction, which must then be preceded by the onset of quantum gravity.  

Our result extends these previous results to the context of a nearly Schwarzschild black hole, spinning only due to its quantum fluctuations. Of course, such a well isolated black hole shielded from any sort of accretion is an idealization.
However, we find it quite remarkable that this highly isolated system has a spacelike onset of quantum gravity, so that even quantum fluctuations do not invalidate this feature of quantum black holes. The complete consistency of all of these results leads to a remarkable principle: \textit{the onset of quantum gravity is always spacelike}. This principle may be useful in more deeply characterizing quantum gravity in the strong gravity regime. 

There is also growing interest in what happens after the onset of quantum gravity, including: evolution through the quantum region leading to a white hole remnant scenario \cite{BianchiEtAl:2018};   quantum bounces \cite{OlmedoSainiSingh:2017}; Planck stars \cite{RovelliVidotto:2014}; and the gauge-gravity no transmission principle, considered for instance in \cite{Engelhardt:2016}. These works have focused on the effects of spacelike singularities. Quantum gravity is also expected to resolve the singularities of rotating black holes, but no mechanism for the timelike case has been identified. Our results show that the emphasis on spacelike transitions is appropriate, at least for small angular momenta. In the presence of more extreme back reaction additional tools will need to be developed. 

The microcanonical ensemble of Eq.~\eqref{eq:PMJ} applies to primordial black holes. While the effects of the quantum fluctuations of spin on the inside of a black hole cannot be probed via astrophysical measurements,\footnote{This is forbidden by the event horizon, but event horizons are teleological and this statement may need to be weakened for dynamical horizons.} the distribution $P_M(\vec{J})$, and through it the assumption (i), could be constrained by LIGO, Virgo, and future gravitational wave observatories. A population of primordial black holes with spin prior determined by \eqref{eq:PMJ} and allowed to evolve through a few generations of mergers could give rise to a distribution of spins in present day black holes distinct from the evolutions of the uniform prior considered in \cite{GerosaBerti:2017}. If so, the imminent transition  from single gravitational wave events \cite{AbbottEtAl:2016} to population analyses could probe the assumption (i), the consequences of this assumption discussed here, and provide the first experimental test of the Bekenstein-Hawking entropy.

\section*{Acknowledgements}
HMH thanks the IGC at Pennsylvania State University for warm hospitality and support while beginning this work and the Perimeter Institute for Theoretical Physics for generous sabbatical support.  This work is supported by Perimeter Institute for Theoretical Physics. Research at Perimeter Institute is supported by the Government of Canada through Industry Canada and by the Province of Ontario through the Ministry of Research and Innovation.

\vfil

\end{document}